\documentclass[sigconf]{acmart}

\copyrightyear{2025}
\acmYear{2025}
\setcopyright{cc}
\setcctype{by}
\acmConference[SC Workshops '25]{Workshops of the International Conference for High Performance Computing, Networking, Storage and Analysis}{November 16--21, 2025}{St Louis, MO, USA}
\acmBooktitle{Workshops of the International Conference for High Performance Computing, Networking, Storage and Analysis (SC Workshops '25), November 16--21, 2025, St Louis, MO, USA}
\acmDOI{10.1145/3731599.3767587}
\acmISBN{979-8-4007-1871-7/2025/11}

\usepackage{booktabs}
\usepackage{float}
\usepackage{graphicx}
\usepackage{ifpdf}
\usepackage[utf8]{inputenc}
\usepackage{listings}
\usepackage{multirow}
\usepackage{url}
\usepackage{textcomp}
\usepackage{tikz}
\usepackage{xcolor}
\usepackage{xspace}
\usepackage{comment}
\usepackage[nameinlink]{cleveref}
\usepackage{subcaption}

\Crefname{figure}{Fig.}{Figs.}
\newif\ifdraft{}

\ifdraft{}
  \newcommand{\mtnote}[1]{ {\textcolor{orange} { ***matteo: #1 }}}
  \newcommand{\jhanote}[1]{ {\textcolor{red} { ***shantenu: #1 }}}
  \newcommand{\miknote}[1]{ {\textcolor{brown} { ***mikhail: #1 }}}
  \newcommand{\amnote}[1]{ {\textcolor{blue} { ***andre: #1 }}}
  \newcommand{\ooknote}[1]{ {\textcolor{blue} { ***ozgur: #1 }}}
  \newcommand{\twnote}[1]{ {\textcolor{blue} { ***tianle: #1 }}}
  \newcommand{\generalnote}[1]{ {\textcolor{gray} { *note: #1 }}}

\else
  \newcommand{\mtnote}[1]{}
  \newcommand{\jhanote}[1]{}
  \newcommand{\miknote}[1]{}
  \newcommand{\amnote}[1]{}
  \newcommand{\ooknote}[1]{}
  \newcommand{\twnote}[1]{}
  \newcommand{\generalnote}[1]{}
\fi

\newcommand{\UP}{\vspace*{-1.0em}}

\newcommand{\B}[1]{\textbf{#1}\xspace}

\ifpdf{}
  \DeclareGraphicsExtensions{.pdf, .jpg, .tif}
\else
  \DeclareGraphicsExtensions{.eps, .jpg, .ps}
\fi

\newlength\myheight
\newcommand*\circled[1]{\settowidth{\myheight}{#1}%
    \raisebox{-.1\myheight}{\tikz[baseline=(char.base)]{%
        \node[shape=circle,draw,minimum size=\myheight*\myheight*.4,inner sep=1pt](char){#1};}}}

\begin{document}
\pagestyle{plain}
\title{Integrating and Characterizing HPC Task Runtime Systems for hybrid AI-HPC workloads}

\author{Andre Merzky}
\affiliation{%
  \institution{RADICAL-Computing Inc.}
  \city{Wilmington, DE}
  \country{USA}
}
\email{andre@radical-consulting.com}

\author{Mikhail Titov}
\affiliation{%
  \institution{Brookhaven National Laboratory}
  \city{Upton}
  \state{NY}
  \country{USA}
}
\email{mtitov@bnl.gov}

\author{Matteo Turilli}
\affiliation{%
  \institution{Rutgers University -- New Brunswick}
  \city{New Brunswick}
  \state{NJ}
  \country{USA}
}
\affiliation{%
  \institution{IE University}
  \city{Madrid}
  \country{Spain}
}
\email{matteo.turilli@rutgers.edu}

\author{Shantenu Jha}
\affiliation{%
  \institution{Rutgers University -- New Brunswick}
  \city{New Brunswick}
  \state{NJ}
  \country{USA}
}
\affiliation{%
  \institution{Princeton Plasma Physics Laboratory\\Princeton University }
  \city{Princeton}
  \state{NJ}
  \country{USA}
}
\email{shantenujha@acm.org}

\begin{abstract}
Scientific workflows increasingly involve both HPC and machine-learning tasks, combining MPI-based simulations, training, and inference in a single execution. Launchers such as Slurm’s \texttt{srun} constrain concurrency and throughput, making them unsuitable for dynamic and heterogeneous workloads. We present a performance study of RADICAL-Pilot (RP) integrated with Flux and Dragon, two complementary runtime systems that enable hierarchical resource management and high-throughput function execution. Using synthetic and production-scale workloads on Frontier, we characterize the task execution properties of RP across runtime configurations. RP+Flux sustains up to 930~tasks/s, and RP+Flux+Dragon exceeds~1,500 tasks/s with over 99.6\% utilization. In contrast, \texttt{srun} peaks at 152~tasks/s and degrades with scale, with utilization below 50\%. For IMPECCABLE.v2 drug discovery campaign, RP+Flux reduces makespan by 30--60\% relative to \texttt{srun}/Slurm and increases throughput more than four times on up to 1,024. These results demonstrate hybrid runtime integration in RP as a scalable approach for hybrid AI-HPC workloads.

\end{abstract}



\maketitle

\section{Introduction}\label{sec:intro}

The scale and complexity of scientific applications executed on high-performance computing (HPC) platforms have evolved from monolithic MPI jobs to heterogeneous workflows that incorporate simulation, machine learning (ML), and data analytics. These workflow applications combine components with diverse computational and resource characteristics into multi-stage, nested workloads, each with its own traditional task dependencies. This evolution introduces significant heterogeneity in execution models, runtime behaviors, and resource requirements, ranging from tightly coupled MPI tasks to short-lived, stateless Python functions, and from long-duration training ML tasks on multi-GPU nodes to bursts of high-throughput, concurrent inference tasks. 

Leadership-class HPC systems provide the raw capability to handle complex workloads at scale. However, these capabilities are often underutilized without task execution runtime systems that can adapt dynamically to the diversity of modern workloads. Effective task execution runtime support must address differences in task granularity, concurrency needs, coupling patterns, and execution environments. In practice, no single runtime or resource management system can efficiently manage this heterogeneity. Modern workflows, however, require combining multiple runtime systems, each tailored for a specific execution model.

System-level resource and job management systems (RJMS) such as Slurm, PBS, and LSF were designed for coarse-grained, static workloads and offer limited support for fine-grained, adaptive, or high-throughput execution. These limitations render them poorly suited for managing workflows comprising thousands of heterogeneous tasks with dynamically changing control flow and asynchronous dependencies. Challenges include low task launch throughput, restricted concurrency in job execution, and the inability to efficiently co-schedule tightly coupled or nested workflows.

To address these challenges, we developed RADICAL-Pilot (RP), a pilot-enabled runtime system that separates resource acquisition from task execution, supporting multi-level scheduling, late binding, and adaptive execution~\cite{merzky2021design}. RP is designed as a modular middleware component rather than a monolithic workflow engine. Its building-block architecture~\cite{turilli2019middleware} emphasizes extensibility and enables integration with third-party systems to support a wide range of execution models. RP is used by both general-purpose workflow systems~\cite{babuji2019parsl} and domain-specific frameworks~\cite{hruska2020extensible,dakka2018high}, serving as the runtime foundation connecting high-level workflow logic with the operational needs of executing diverse workloads on HPC platforms.

Multi-stage, multi-component workflows—such as high-\allowbreak through\-put virtual screening exemplified by the IMPECCABLE application~\cite{saadi2020impeccable,lee2020scalable}—capture the needs of modern HPC applications. IMPECCABLE includes diverse, heterogeneous, and adaptive workloads, ranging from GPU-bound ML model training to CPU-intensive molecular simulations and MPI-based scoring calculations. Each workload features unique concurrency, duration, and data dependencies. 

The current generation of IMPECCABLE (i.e., IMPECCABLE.v2, hereafter, IMPECCABLE) requires the simultaneous execution of independent tasks (both executables and functions), adaptive scheduling based on runtime feedback, and efficient utilization of heterogeneous hardware across thousands of nodes. The variety of tasks demands different execution models: lightweight launchers for high-throughput screening and surrogate inference, hierarchical scheduling for co-scheduling physics-based scoring, and runtime environments capable of managing dynamic resource allocation for asynchronous model training and ensemble simulations. No single runtime can meet these needs, making it essential to combine multiple specialized runtimes within a unified middleware. 

In this paper, we extend the work previously done to integrate RP with Flux, extending supported capabilities, dramatically improving performance and adding the concurrent integration of Dragon~\cite{lee2021scalable,turilli2024exaworks,al2021exaworks}.

Our contributions are threefold. First, we extend RP to integrate with Flux and Dragon, enabling adaptive and high-throughput execution on Frontier, a DOE leadership-class exascale platform. Second, we analyze RP’s performance under different workload configurations, focusing on throughput, resource utilization, and runtime overhead. Third, we test the system using both synthetic benchmarks and a production-scale IMPECCABLE workflow, showcasing the benefits of combining specialized runtimes within a single, unified middleware framework.

Across seven experiments, we demonstrate that RP can sustain high task throughput, maintain near-perfect resource utilization, and scale across thousands of nodes on Frontier. Using Flux as a backend for executing independent executable tasks (i.e., not functions), RP achieves up to 744~tasks/s with a single Flux instance and over 900~tasks/s with multiple Flux instances on 1024 nodes. A hybrid configuration with Flux and Dragon concurrently executing both executable and function tasks exceeds 1,500~tasks/s while maintaining 99.8--100\% utilization. In contrast, RP with Slurm’s \texttt{srun} is limited by system-level concurrency ceilings, capping utilization at 50\%. When applied to the IMPECCABLE workflow, RP with Flux shortens makespan by 30--60\% relative to Slurm and sustains high concurrency across 256--1024 nodes. These results confirm RP’s ability to orchestrate heterogeneous, adaptive workloads at scale while mitigating the constraints of traditional launchers.

\section{Exemplar Use Case}\label{sec:uc}

IMPECCABLE is a drug discovery computational campaign (i.e., workflow of workflows) that exemplifies the need for runtime systems capable of handling workloads with highly variable, heterogeneous resource requirements. It combines machine learning with physics-based methods into a set of interdependent workflows, each with different execution models and compute setups. This variety makes IMPECCABLE representative of a broader range of scientific workflows whose complexity exceeds what traditional, monolithic runtime systems can manage.

The IMPECCABLE campaign  includes six main workflows: (1) high-throughput molecular docking (CPU-only, up to 128 nodes); (2) training a Simple SMILES Transformer (SST) surrogate model (GPU, up to 4 nodes); (3) large-scale SST-based surrogate inference (GPU, up to 128 nodes); (4) physics-based scoring with Dock-Min-MMPBSA and molecular property prediction using AMPL (CPU/GPU, up to 128 and 16 nodes, respectively); (5) ensemble simulations with ESMACS (CPU/GPU, up to 625 nodes); and (6) de novo molecule generation with REINVENT (GPU, 1 node). These workflows encompass various task types: some are embarrassingly parallel, or require multi-node MPI coupling, and still depend on frequent feedback between the learning and sampling stages.

Beyond resource diversity, IMPECCABLE sub-workflows differ significantly in how tasks are implemented. Physics-based simulations and molecular docking stages are usually launched as standalone executables—compiled binaries (e.g., AMBER, AutoDock) that require dedicated system processes and, in some cases, MPI runtime coordination. In contrast, the machine learning components—such as SST training and inference, or the REINVENT generative model—are often implemented as Python functions or scripts, called within long-running Python processes or through dynamic scheduling backends. These Python-based workloads often rely on in-process memory sharing, GPU context reuse, and tighter feedback loops, which are challenging to replicate with executables. Supporting both execution styles—standalone binaries and Python functions—within the same workflow demands a runtime system capable of managing differences in task launch modes, environment isolation, data movement, and process lifecycle management.

The campaign also spans a range of coupling models. Certain stages, such as high-throughput docking and surrogate inference, consist of thousands of loosely connected tasks that can run independently with minimal coordination, making them ideal for high-throughput, low-latency launchers. Other stages, like ensemble simulations and MPI-based scoring, involve tightly coupled tasks that must be launched concurrently with co-scheduled resources and coordinated runtime behavior. There are also intermediate forms of coupling. For example, REINVENT-based generation and SST-guided patch selection involve asynchronous pipelines of Python functions communicating through in-memory data structures or message queues. These data-dependent tasks rely on fast, low-overhead communication across processes or nodes, making them unsuitable for either completely decoupled or tightly synchronized execution models.

These distinctions in task implementation and coupling place strict requirements on the runtime environment that no single system can fully meet. High-throughput docking and inference workloads benefit from flat, low-overhead launchers that can cycle through thousands of tasks per minute. Physics-based scoring stages require hierarchical scheduling and coordinated multi-node resource placement. Intermediate cases benefit from shared memory abstractions and lightweight coordination primitives that enable high-throughput yet stateful execution. As such, IMPECCABLE is not only a demanding computational campaign, but also a compelling case for integrating multiple specialized runtimes—each optimized for different execution models—within a unified middleware like RADICAL-Pilot.

Emerging use cases further amplify this need. Reinforcement learning agents, active learning loops, and streaming pipelines introduce execution behaviors that blur the line between tightly and loosely coupled tasks. These workflows often require persistent services (e.g., learners, replay buffers), dynamic spawning of short-lived workers, and rapid data exchange without blocking synchronization. Supporting such patterns at scale demands runtime systems that provide both high launch throughput and lightweight coordination mechanisms—capabilities that lie outside the scope of traditional batch schedulers or MPI-based systems. IMPECCABLE thus anticipates a broader class of adaptive and data-driven scientific workflows, making it a relevant benchmark for evaluating future middleware designs.

\section{Design and Implementation}\label{sec:design}

RADICAL-Pilot (RP) is a pilot system~\cite{turilli2018comprehensive} designed to support the execution of heterogeneous workloads on leadership-class HPC platforms. RP implements the pilot abstraction, which decouples resource acquisition from task execution through multi-level scheduling and late-binding mechanisms. This design enables RP to manage dynamic workloads that include simulation, data processing, machine learning, and other task types with varying granularity, duration, and coordination requirements.

RP implements a building-block architectural approach~\cite{turilli2019middleware}, emphasizing self-sufficiency, interoperability, composability, and extensibility. These principles enable modular integration with third-party systems and facilitate scalable composition of workflow infrastructures. Over the past decade, RP has been used in multiple DOE and NSF projects, integrated with workflow engines such as Parsl~\cite{alsaadi2022radical}, Swift/T~\cite{turilli2016integrating}, and PanDA~\cite{merzky2019panda}, and with distributed computing frameworks like Apache Spark~\cite{paraskevakos2018task} and runtime systems including PRRTE~\cite{al2021exaworks}.

At the core of RP are two abstractions: pilots, which represent resource placeholders, and tasks, which represent individual units of work. Each abstraction is modeled through a state machine and coordinated via an event-driven execution engine. This design supports a wide range of execution patterns, e.g., high-throughput ensembles, multi-node MPI applications, standalone executables, and Python functions. This makes RP an effective runtime substrate for supporting ensemble-based simulations, deep-learning-driven adaptive molecular dynamics simulations, and active/reinforcement-learning methods for scientific applications.

In this work, we focus on RP's Agent, which acquires resources and manages task execution. We extended the Agent to concurrently instantiate and coordinate multiple task runtime systems, enabling selection of the most suitable execution model. This adaptive mapping is essential for complex, multi-paradigm workflows like IMPECCABLE, where diverse workloads benefit from runtimes with distinct strengths (e.g., high-throughput launchers, hierarchical schedulers). By making these choices dynamically while preserving a unified middleware interface, RP enables workflow systems to leverage the best features of diverse task runtime systems without incurring integration complexity.

Note that, by execution model, we mean the way tasks are scheduled, placed, launched, and coordinated on a given allocation, capturing their granularity, concurrency, coupling, and runtime environment. For example, tightly coupled MPI jobs, GPU-bound training runs, and fine-grained stateless functions each require distinct execution models, differing in launch method, resource binding, and communication patterns.

\subsection{Supporting HPC leadership-class platforms}\label{ssec:leadership}

Leadership-class platforms like Frontier present operational challenges that stress RP's design through bespoke architectures, strict resource management policies, and platform-specific constraints. Traditional MPI-based launch mechanisms suffer from high startup latencies, centralized bottlenecks, and limited user-space control, necessitating alternative execution models decoupled from system-level schedulers.

These platform constraints intersect with workflow-level diversity in task implementation and coupling patterns. Tasks may be standalone executables (compiled binaries, MPI applications) or Python functions (ML, analytics workloads), each requiring different execution environments. Coupling patterns---loosely coupled (independent tasks), tightly coupled (coordinated multi-node execution), and data-coupled (shared memory communication)---further determine runtime requirements. Loosely coupled tasks benefit from low-latency, high-throughput launchers; tightly coupled workloads require co-scheduling and placement control; data-coupled tasks need coordination mechanisms without full synchronization.

To address this diversity, we extended RP to incorporate two complementary runtime systems within a single resource allocation. Flux offers hierarchical, policy-driven scheduling with fine-grained placement control, making it ideal for executable tasks that require coordination and locality. Dragon offers lightweight, high-throughput dispatch optimized for Python functions and asynchronous workloads, using shared-memory coordination and process pooling for minimal overhead.

This dual-backend approach enables RP to dynamically route tasks based on their properties and coupling requirements, adapting to both platform constraints and application diversity while maintaining a unified middleware interface. Tasks are mapped to the backend that best matches their execution models: multi-node or tightly coupled tasks to Flux, short CPU-bound or Python function tasks to Dragon.

\subsection{Integration with Flux and Dragon}\label{ssec:flux-dragon}

\begin{figure}
  \centering
  \includegraphics[trim=0 3 0 1,clip,width=0.46\textwidth]{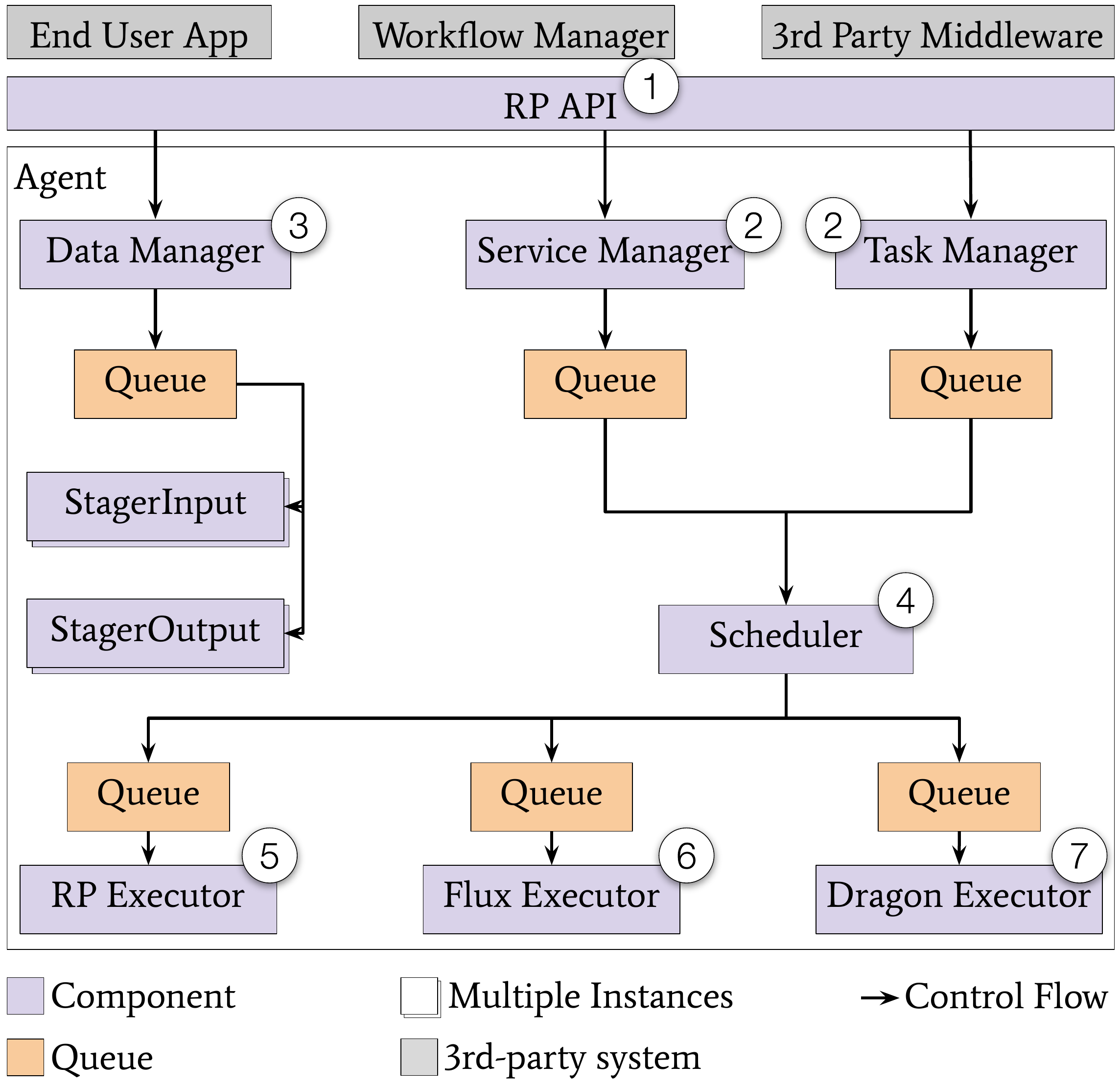}
    \caption{Architecture of RADICAL-Pilot's integration with task runtime systems, such as Flux and Dragon. Stacked boxes (e.g., StagerInput, StagerOutput) indicate that multiple instances of that component can execute concurrently. }\label{fig:flux-dragon}
\end{figure}

RP integrates with both Flux and Dragon by extending its Agent component through specialized execution subsystems. Fig.~\ref{fig:flux-dragon} illustrates the high-level integration of RP with Flux and Dragon. RP receives pilot, task, or service descriptions via its API directly from a user-defined application, a workflow manager, or any other 3rd party middleware that loads RP's Python API (Fig.~\ref{fig:flux-dragon}~\circled{1}). After a pilot becomes active, RP starts to manage the given task and, in case, service descriptions (Fig.~\ref{fig:flux-dragon}~\circled{2}), staging the required input data (Fig.~\ref{fig:flux-dragon}~\circled{3}) and scheduling the task and service for execution (Fig.~\ref{fig:flux-dragon}~\circled{4}). RP's scheduler can queue the tasks on the RP's executor (Fig.~\ref{fig:flux-dragon}~\circled{5}) or on a 3rd party low-level runtime system as Flux or Dragon (Fig.~\ref{fig:flux-dragon}~\circled{6},\circled{7}).

In both cases, RP's modular architecture ensures that the Flux and Dragon integrations are cleanly isolated within the Agent's launching and executing subsystems. This design enables dynamic configuration of the Agent to select either Flux or Dragon depending on workload and resource characteristics, without requiring changes to the broader RP runtime or API. As a result, RP can flexibly optimize for different workflow requirements: using Flux for adaptive, or loosely coupled workflows that require nested scheduling and resource heterogeneity, and Dragon for homogeneous, high-throughput workloads where launch performance is critical.

The integration with Flux and Dragon also preserves RP's capability to manage input and output staging, task pre- and post-processing, and failure recovery uniformly across different execution substrates. Tasks launched via Flux or Dragon continue to pass through RP's full task lifecycle, ensuring consistent behavior and enabling RP to provide comprehensive profiling, tracing, and runtime analytics regardless of the underlying launcher.

\subsubsection{Flux integration}

\begin{figure}
  \centering
  \includegraphics[trim=0 3 0 1,clip,width=0.29\textwidth]{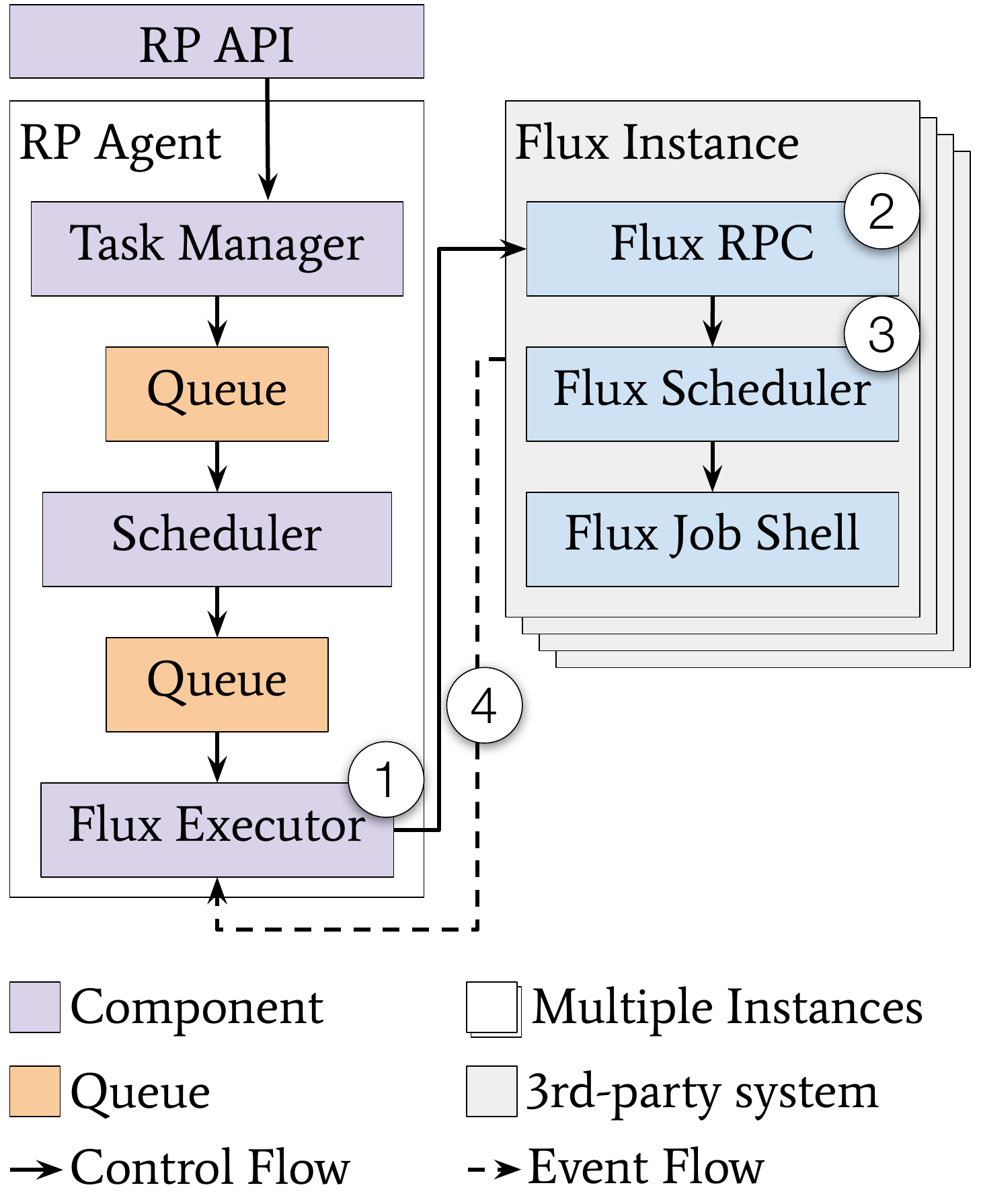}
    \caption{Architecture diagram of the integration between RADICAL-Pilot and Flux. RP's Flux executor can concurrently drive multiple Flux instances to scale execution and throughput. RP and Flux are fully integrated at the control and event level.}\label{fig:rp-flux}
\end{figure}

The integration of Flux into RP's Agent component establishes a scalable, hierarchical execution model that supports heterogeneous workflows. As shown in Fig.~\ref{fig:rp-flux}, within the RP Agent, a dedicated Flux executor is instantiated whenever a workload configuration requests the use of Flux for task execution. This executor interfaces directly with a running Flux instance to submit, monitor, and control tasks.

Task submission proceeds asynchronously: as tasks transition into RP's Flux executor (Fig.~\ref{fig:rp-flux}~\circled{1}), they are serialized into Flux job descriptions and submitted to a Flux instance via the Flux RPC interface (Fig.~\ref{fig:rp-flux}~\circled{2}). Each task specifies resource requirements, e.g., the number of cores, GPUs, memory, and node locality constraints. The Flux scheduler, running inside the instance, manages fine-grained resource placement across CPUs, GPUs, and other resources, applying scheduling policies such as first-come-first-served, backfilling, or customized co-scheduling strategies (Fig.~\ref{fig:rp-flux}~\circled{3}).

Task state transitions are coordinated through event-driven communication between RP and Flux (Fig.~\ref{fig:rp-flux}~\circled{4}). RP Flux executor subscribes to job lifecycle events emitted by the Flux instance, including job start, completion, failure, and cancellation. These events are consumed asynchronously, updating task states in RP's internal registry. This enables RP to maintain complete consistency of its task state/event model without direct polling or blocking on Flux operations, achieving both high scalability and responsiveness.

Resource management in this integration is explicit and dynamic. Flux partitions the resources allocated to the pilot across tasks according to the runtime scheduling decisions, enabling the concurrent execution of heterogeneous tasks with varying resource footprints. Nested Flux instances and hierarchical scheduling are supported where needed, enabling RP to scale workload execution from a few nodes to thousands of nodes efficiently (see the multiple gray boxes indicating multiple Flux instances in Fig.~\ref{fig:rp-flux}).

Error handling within the Flux integration leverages both RP's and Flux's native capabilities. Failed task submissions, runtime job failures, and communication errors are captured through the Flux event system (Fig.~\ref{fig:rp-flux}~\circled{4}) and mapped into RP's failure handling framework. In case of unexpected Flux daemon failures or API communication issues, RP transitions affected tasks into failure states, logs diagnostic information, and optionally triggers Agent failover or restart procedures depending on the failure mode.

RP's and Flux's logging and tracing are fully integrated. RP's tracing system captures flux job metadata, status changes, and runtime diagnostics that can be correlated with Agent logs to enable comprehensive post-mortem analysis. Through the RADICAL-Analytics profiling capabilities, events such as task submission timestamps, Flux job IDs, and resource assignment details are recorded, supporting the fine-grained characterization of workflow performance.

Flux-RP integration supports advanced workload patterns, including nested workflows, dynamic task generation, adaptive placement, and fault-tolerant execution. It enables RP to sustain high throughput and flexible resource utilization, even on contended HPC platforms where schedulers like Slurm impose throughput or concurrency limits (as on Frontier). Unlike Dragon, which maximizes launch rates with minimal coordination, Flux offers a richer, more adaptive environment suited to workflows with complex resource and control-flow needs.

\subsubsection{Dragon integration}

\begin{figure}
  \centering
  \includegraphics[trim=0 3 0 1,clip,width=0.29\textwidth]{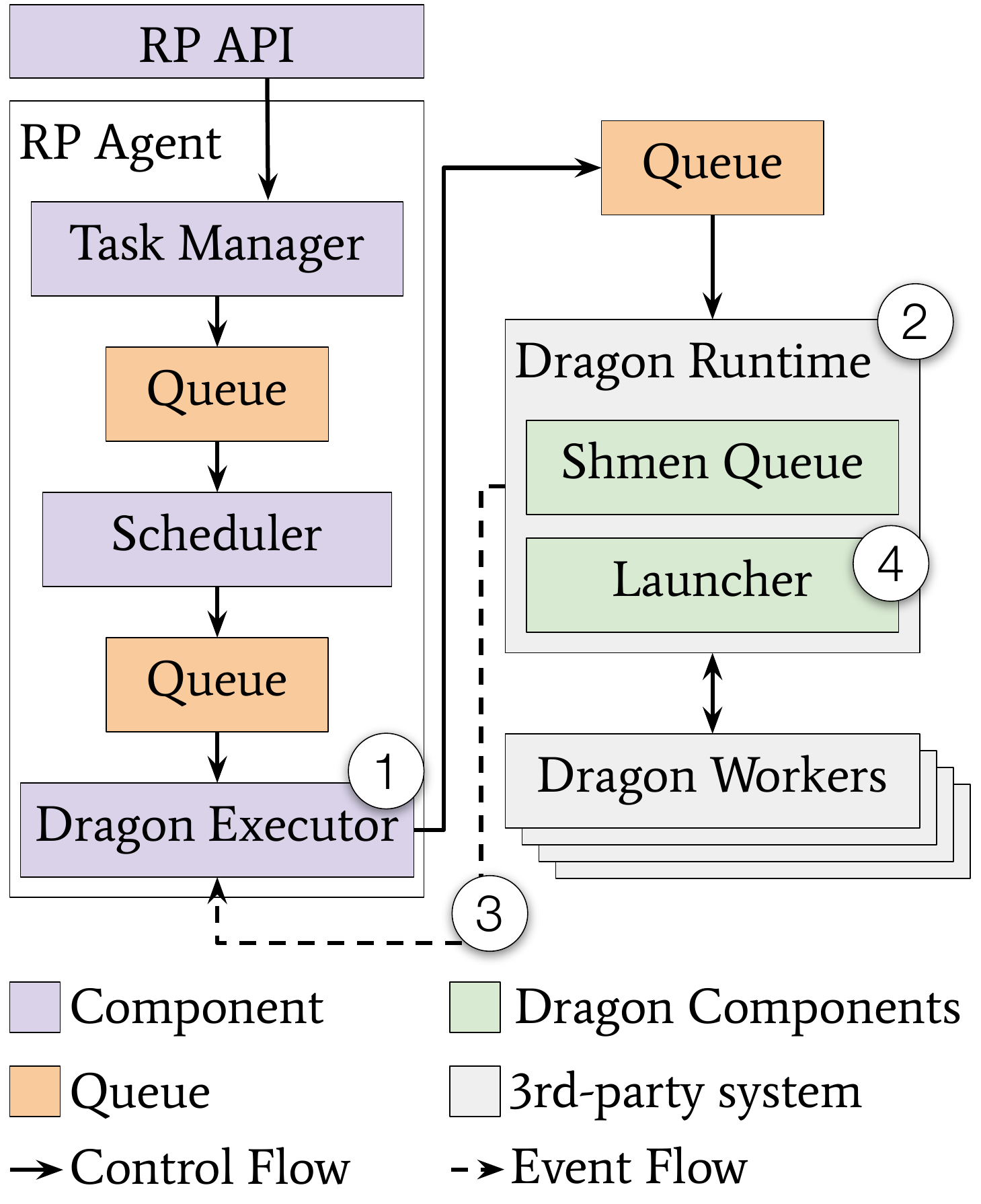}
    \caption{Architecture diagram of the integration between RADICAL-Pilot and Dragon.}\label{fig:rp-dragon}
\end{figure}

The integration with Dragon enables a lightweight, high-throughput executor within RP's Agent. As shown in Fig.~\ref{fig:rp-dragon}, RP's Dragon Launcher instantiates a Dragon runtime multiprocessing environment to manage task execution across the allocated resources (Fig.~\ref{fig:rp-dragon}~\circled{1}). Communication between RP and the Dragon runtime is handled via ZeroMQ pipes: tasks ready for execution are serialized and pushed to the Dragon runtime (Fig.~\ref{fig:rp-dragon}~\circled{2}), while task completion events are asynchronously pushed back to the RP Dragon Executor (Fig.~\ref{fig:rp-dragon}~\circled{3}). A watcher thread receives those events and updates the task states in RP's registry.

Dragon runtime is designed to minimize system overheads: it directly launches tasks on workers without intermediate job scheduling layers, thus achieving high task dispatch rates\amnote{TODO: confirm!} (Fig.~\ref{fig:rp-dragon}~\circled{4}). Resource management within the Dragon runtime is implicit: processes are spawned into the resource environment allocated to the RP Agent, without further internal partitioning or explicit co-scheduling. Shared memory (obtained via `Shmem Queue`, see Fig.~\ref{fig:rp-dragon}) among the spawned/launched processes enables internode communication for the tasks that load the application-side Dragon module. Logging from the Dragon runtime is forwarded to RP's logging system via inherited file descriptors, ensuring that execution diagnostics remain consistent with those of RP.

Dragon integration includes robust error-handling. RP monitors the Dragon runtime, detecting failures during startup or execution. If initialization fails or the runtime crashes, RP triggers failover and moves affected tasks to error states. Startup timeouts prevent RP from stalling. These safeguards ensure Dragon's lightweight execution model remains suitable for production without compromising fault isolation or resilience.

In contrast to the Flux integration, which provides fine-grained resource placement, scheduling, and fault-tolerant hierarchical execution, the Dragon integration emphasizes minimal latency, maximal throughput, and operational simplicity. This makes Dragon particularly well-suited for workloads composed of large numbers of short-lived or homogeneous tasks, such as ensemble simulations, parameter sweeps, or lightweight ML inference pipelines. RP's building-block design enables users to select either Flux or Dragon dynamically at runtime, depending on the workload's coordination and throughput requirements, without modifying workflow logic or application interfaces.

\section{Performance Characterization}\label{sec:exp}

We conduct seven experiments to characterize RADICAL-Pilot's (RP) performance with workflows requiring diverse execution models at scale. These span five runtime configurations: RP with Slurm, Flux, Dragon, Flux+Dragon, and a first IMPECCABLE campaign implementation (\S\ref{sec:uc}). Together, they show RP's ability to support scalable, asynchronous, and heterogeneous task execution across varied runtime systems and workload types.

Our performance characterization uses three core metrics: (1) \textit{throughput}, measured as the number of tasks launched per second, independent of their execution duration; (2) \textit{resource utilization}, defined as the percentage of allocated compute resources actively used over time; and (3) \textit{runtime overhead}, representing the infrastructure setup time before workflow execution begins. To systematically explore the design space, we classify experiments along three orthogonal dimensions: execution heterogeneity (varying launchers and task types), resource partitioning (varying number and size of partitions), and task modality (executables or Python functions).

We employ three types of workloads. Null workloads consist of empty tasks that execute no application code and immediately return, minimizing application-level overheads so as to stress only the middleware stack and reveal its internal throughput limits. Dummy workloads execute fixed-duration sleep tasks to emulate sustained system load without performing meaningful computation, keeping execution queues saturated and enabling accurate measurement of peak resource utilization. Finally, the IMPECCABLE campaign is a faithful approximation of the production application, using representative dummy tasks to preserve its heterogeneity, task structure, and execution dynamics. This enables realistic performance evaluation without the cost of full scientific computations.

Table~\ref{tab:experiments} summarizes the experiments across launcher types, pilot sizes, partitioning strategies, and task modalities. The \texttt{srun} experiment uses Slurm on a four-node pilot, launching one-core tasks at full hardware-thread density (4 tasks per core). The \textit{flux\_1} and \textit{flux\_n} experiments evaluate Flux with a single instance scaled up to 1024 nodes or multiple concurrent partitions. \textit{dragon} uses Dragon for executable tasks; \textit{flux+dragon} mixes Flux and Dragon to concurrently launch executables and Python functions. The final two experiments, \textit{impeccable\_srun} and \textit{impeccable\_flux}, run the IMPECCABLE workflow on 256--1024 nodes with up to 1,800 synthetic tasks and varied resource demands (1--7,168 cores, up to 1,024 GPUs), enabling realistic comparisons between Slurm and Flux.

Collectively, these experiments compare RP's performance across task runtime systems, workloads, and resource configurations, demonstrating its ability for high-\allowbreak through\-put, adaptive, heterogeneous execution on leadership-class HPC platforms.

\begin{table*}[ht]
    \caption{Summary of the experiments used to evaluate the performance of RADICAL-Pilot under different execution models, launch mechanisms, and workload configurations. Exec = executable task; func = function tasks; cpn = cores per node.\label{tab:experiments}}
	\centering
    \footnotesize
    \begin{tabular}{lllrrlrr}
	\toprule
    \textbf{Exp ID}            &
    \textbf{Workload}          &
    \textbf{launcher}          &
    \textbf{\#nodes/pilot}     &
    \textbf{\#partitions}      &
    \textbf{task types}        &
    \textbf{\#tasks}           &
    \textbf{\#cores/task}      \\
	%
    \midrule
    \textit{\textbf{srun}}     &
    null, dummy(180s)          &
    srun                       &
    4                          &
    1                          &
    exec                       &
    $n_{nodes} * cpn * 4$      &
    1                          \\
    %
    %
    \textit{\textbf{flux\_1}}  &
    null, dummy(360s)          &
    flux                       &
    1,4,16,64,256,1024         &
    1                          &
    exec                       &
    $n_{nodes} * cpn * 4$      &
    1                          \\
    %
    %
    \textit{\textbf{flux\_n}}  &
    null, dummy(180s)          &
    flux                       &
    64, 1024                   &
    1,4,16,64                  &
    exec                       &
    $n_{nodes} * cpn * 4$      &
    1                          \\
    %
    %
    \textit{\textbf{dragon}}   &
    null, dummy(180s)          &
    dragon                     &
    1,4,16,64                  &
    1                          &
    exec                       &
    $n_{nodes} * cpn * 4$      &
    1                          \\
    %
    %
    \textit{\textbf{flux+dragon}} &
    null, dummy(360s)          &
    flux \& dragon             &
    1,4,16,64                  &
    1                          &
    exec \& funcs              &
    $n_{nodes} * cpn * 4$      &
    1                          \\
    %
    %
    \textit{\textbf{im\-pec\-ca\-ble\_\-srun}}  &
    impeccable                 &
    srun                       &
    256,1024                   &
    1                          &
    exec                       &
    $\sim$550,$\sim$1800       &
    1-7168                     \\
    %
    %
    \textit{\textbf{im\-pec\-ca\-ble\_\-flux}}  &
    impleccable                &
    flux                       &
    256,1024                   &
    1                          &
    exec                       &
    $\sim$550,$\sim$1800       &
    1-7168                     \\
    %
    %
    %
    %
	\bottomrule
	\end{tabular}
    \UP
\end{table*}

\subsection{Scaling Runtime Task Capabilities}\label{ssec:exp-runtime}

To characterize the performance of RP with different task runtime systems used as backend executors, we conduct a sequence of experiments using \texttt{srun}, Flux, Dragon, and Flux and Dragon combined. These experiments isolate key architectural trade-offs in how each backend handles task launching, concurrency, partitioning, and runtime overheads. We evaluate each configuration by measuring task throughput (in tasks per second) and resource utilization (as a percentage of available cores actively used—note that using GPUs would lead to the same results as we are executing dummy workloads, and placement/launching performance is independent of the type of computing support utilized). Together, these results establish a baseline for understanding how the integration of runtime systems affects RP's scalability and efficiency.

Experiment~\textit{srun} measures RP's performance when using the native \texttt{srun} command on Frontier to launch tasks. As \texttt{srun} is subject to system-level constraints that limit the number of concurrent invocations within a user allocation, this experiment exposes the bottlenecks introduced by such limitations. In particular, Frontier enforces a ceiling on concurrent \texttt{srun} executions, capping task-level concurrency regardless of the number of available compute nodes. As a result, task throughput is significantly constrained and resource utilization drops, especially at larger node counts. This experiment provides a baseline for comparing more flexible and scalable launching mechanisms.

Experiment~\textit{flux} evaluates RP with Flux. Flux supports hierarchical scheduling and dynamic resource management within a job allocation, allowing multiple Flux instances to run concurrently on disjoint resource partitions. RP leverages this capability by spawning multiple sub-agents, each managing a local Flux instance and its own partition. This decentralized design improves task launch concurrency, mitigates scheduling contention, and scales more effectively with node count. The experiment demonstrates that integrating Flux with RP achieves higher task throughput and better resource utilization, particularly at scale, where centralized launchers like \texttt{srun} become performance bottlenecks.

Experiment \textit{dragon} investigates the performance of RP and Dra\-gon. Dragon is designed for the efficient execution of in-memory Python functions and lightweight message-passing between processes. However, for comparison with other backends, we configure Dragon to launch external executable tasks. In this setup, a single Dragon instance spans the entire allocation. Unlike Flux, Dragon does not yet support partitioning the allocation into independently managed subdomains. As such, the experiment highlights Dragon's ability to sustain high launch rates in smaller allocations, while also revealing scalability challenges under centralized resource control at larger scales. Future work will investigate partitioned configurations using Dragon to enable concurrency and resilience similar to our approach with Flux.

Experiment~\textit{flux+dragon} explores RP's ability to manage heterogeneous workloads by simultaneously deploying Flux and Dragon. In this configuration, RP uses Flux to launch executable tasks and Dragon to launch Python function tasks, with each runtime deployed in multiple concurrent partitions. This hybrid setup shows RP's flexible resource partitioning and runtime orchestration capabilities, enabling task-type-aware backend selection, high concurrency, and resilience to failure. The experiment measures task throughput and resource utilization when executing a mixed workload across 2--64 nodes with varying numbers of Flux and Dragon partitions. The results show that combining multiple task runtime systems within RP increases throughput while preserving high resource utilization and runtime efficiency.

\subsubsection{Experiment \textit{Srun}}\label{sssec:srun}

By default, Frontier allows running multiple \texttt{srun} commands concurrently. We can use a single \texttt{srun} to place and launch each task, specifying the number and type of resources needed as command arguments. It is important to note that the number of concurrent \texttt{srun} commands that can be issued, along with their execution queue depth in Slurm, is determined by Frontier's system administrators.

We perform Experiment~\textit{srun} (see Table~\ref{tab:experiments}) to quantify the maximum task concurrency and throughput achievable under these constraints. We run a synthetic workload composed of independent, single-core tasks, each of which sleeps for 180 seconds. This workload is distributed across 4 compute nodes, each configured with SMT=1 for a total of 224 cores.

\begin{figure}
    \includegraphics[width=0.5\textwidth]{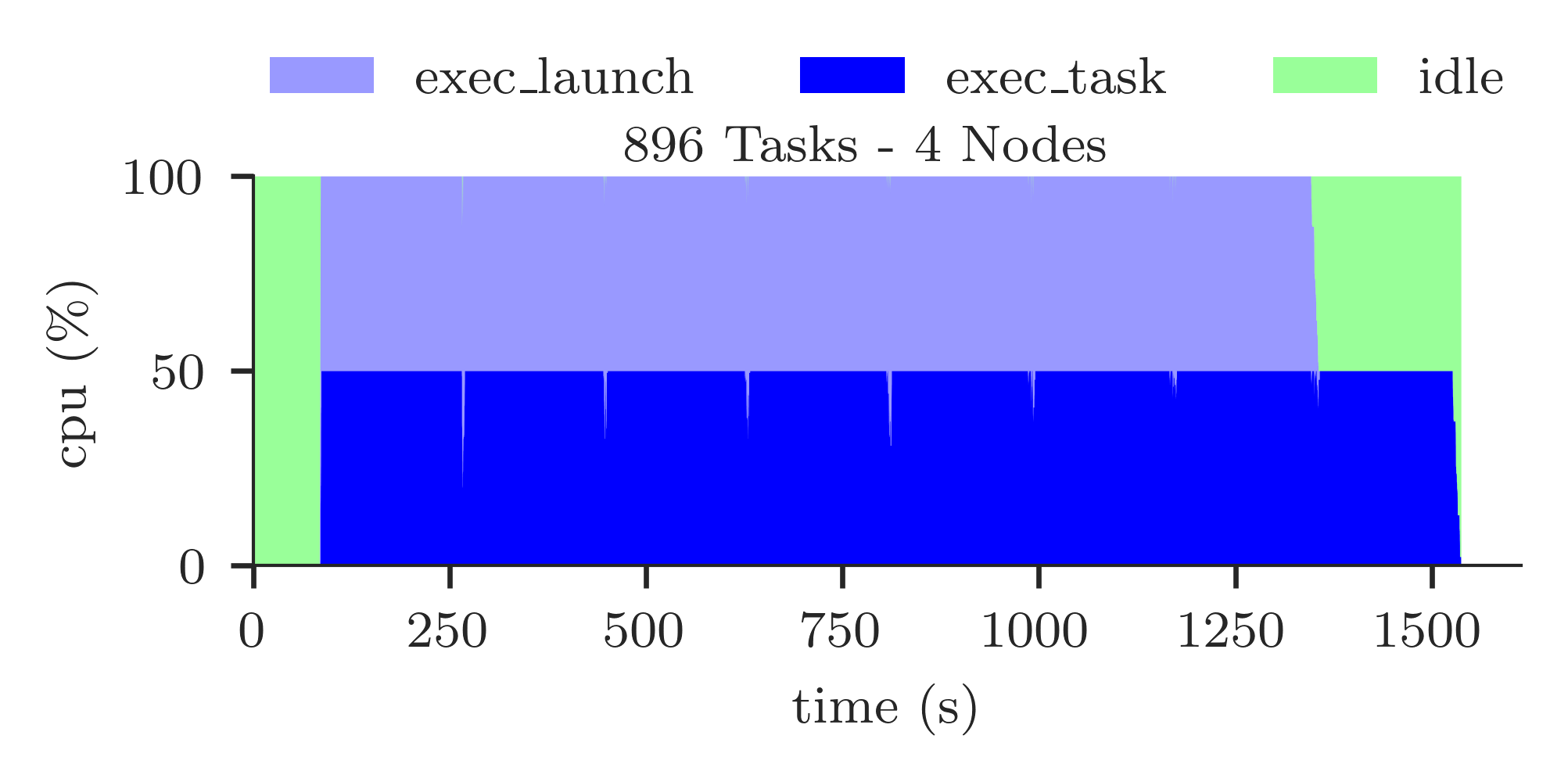}
    \caption{\B{Srun Resource utilization:} Execution of 896 single-core dummy 180s-long tasks on 4 nodes. Frontier's Srun concurrency ceiling limits resource utilization to 50\%.
    \label{fig:srun_concurrency}}
\end{figure}

As shown in Fig.~\ref{fig:srun_concurrency}, we observe a maximum concurrency of 112 tasks, i.e., using only half of the available cores. This limit remains consistent across different node counts, indicating that Frontier enforces a system-wide cap on the number of concurrently active \texttt{srun} processes. As a result, resource utilization does not increase with larger workloads, leading to proportionally greater resource underutilization in bigger jobs and workloads. Fig.~\ref{fig:all_throughput}(a) confirms these results, showing how Slurm's task throughput decreases with the number of nodes.

This experiment reveals a key limitation of relying only on Slurm's native task launching via \texttt{srun} for workloads that require high concurrency. In workflows like IMPECCABLE, which need dynamic, fine-grained, and high-\allowbreak through\-put heterogeneous task execution, this restriction greatly reduces overall efficiency. To address this bottleneck, we turn to Flux and Dragon runtime systems that, as seen in \S\ref{sec:design}, separate task launching from Slurm.

\subsubsection{Experiment \textit{Flux\_1}}

In this experiment, we characterize the performance of RP when integrated with a single Flux instance. We measure the task throughput capabilities of a single Flux instance launched and managed by RP across increasing node counts. As in the previous experiment, we use a synthetic workload composed of single-core, 180s sleep tasks to isolate launching and scheduling overheads from task execution costs.

\begin{figure*}
    \includegraphics[width=1\textwidth]{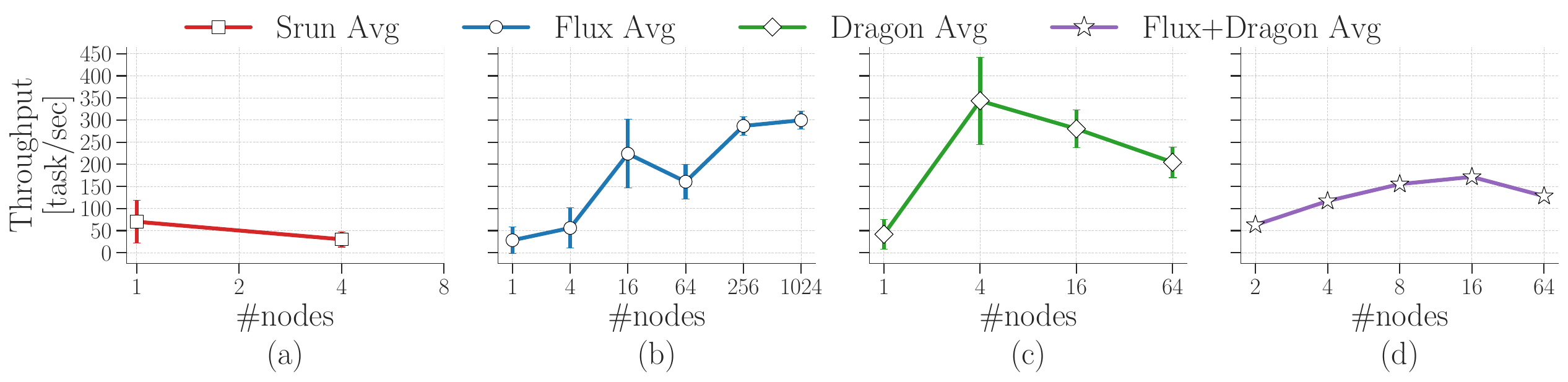}
    \caption{Average (avg) task throughput for Srun (a), Flux (b), Dragon (c), and Flux+Dragon (d) runtime systems, executing from 224 to 229,376 single-core tasks on 1 to 1024 nodes. Comparatively, all the runtime systems have similar performance with one node, but Slurm throughput already decreases with two nodes. Flux throughput mostly scales with node count, while Dragon shows a constant high throughput independent of node count.
    \label{fig:all_throughput}}
\end{figure*}

Fig.~\ref{fig:all_throughput}(b) shows the average throughput rates for jobs ranging from 1 to 1024 nodes. We observe a strong positive correlation between node count and task throughput, with average throughput increasing from approximately 28~tasks/s at 1 node to nearly 300~tasks/s at 1024 nodes. Peak throughput reaches 744~tasks/s, highlighting the efficiency of Flux's internal task scheduling and launching mechanisms when operating at scale.

Notably, even a single Flux instance sustains high throughput across all scales tested, and its performance continues to improve with larger resource partitions. This scaling behavior contrasts with the \texttt{srun} experiment (Fig.~\ref{fig:all_throughput}(a)), where task throughput degrades sharply beyond a single node.

We also observe substantial throughput variability across repetitions at each scale. This variability may reflect the sensitivity of Flux performance to background system load, variability in RPC latency, or internal queueing dynamics within Flux. Nonetheless, the average rates remain consistently high, enabling Flux to maintain high resource utilization across a wide range of node counts.

The results of this experiment confirm that RP+Flux constitutes an effective stack for high-\allowbreak through\-put execution of executable tasks at scale, overcoming the concurrency bottlenecks imposed by \texttt{srun}-based task launching.

\subsubsection{Experiment \textit{Flux\_n}}

In this experiment, we increase the number of concurrently running Flux instances from 1 to 64 to evaluate the impact of partitioning on Flux's task throughput and scalability. Each instance manages a unique subset of compute nodes, forming a dedicated partition. This experiment builds on the results of \textit{flux\_1}, determining whether distributing the workload across multiple Flux instances can increase throughput and/or improve utilization.

Beyond performance, multiple instances enhance fault tolerance by isolating failures: if a node or instance crashes, only the corresponding instance is affected. However, using multiple instances increases coordination overhead and introduces additional complexity. Moreover, because each Flux instance is launched via \texttt{srun}, this experiment is subject to Frontier's limit of 112 concurrent \texttt{srun} invocations measured in \S\ref{sssec:srun}. 

\begin{figure*}
    \includegraphics[width=1\textwidth]{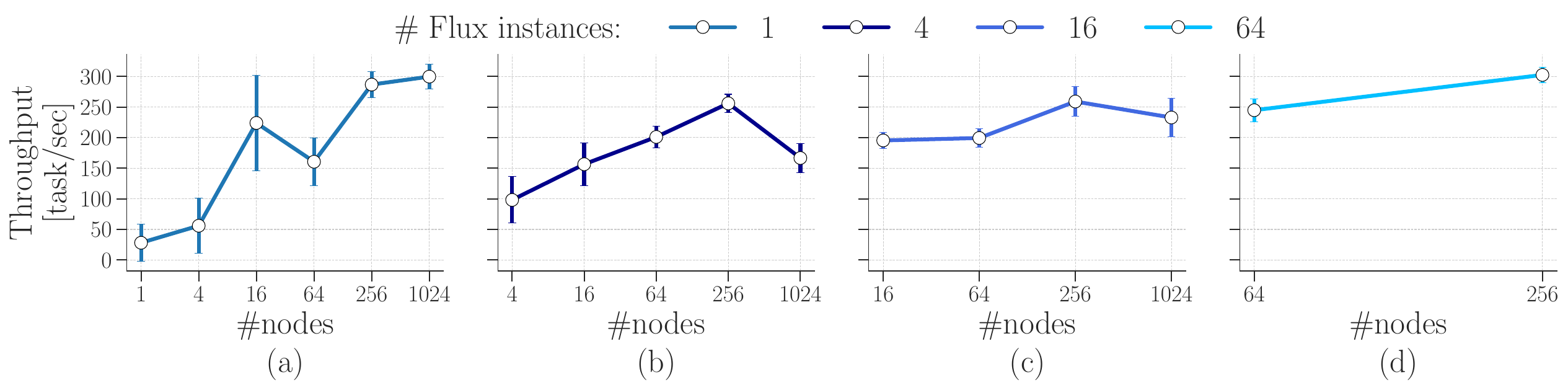}
    \caption{Average Flux throughput with between 1 and 4 concurrent instances, executing from 224 to 229,376 single-core tasks on 1 to 1024 nodes. Throughput scales with node count and with the number of flux instances.
    \label{fig:flux_n_throughput}}
\end{figure*}

Fig.~\ref{fig:flux_n_throughput} shows the average throughput across various node and instance counts. For small jobs (e.g., 4 or 16 nodes), increasing the number of instances improves throughput. For instance, with 4 nodes, increasing from 1 to 4 instances raises average throughput from 56 to 98~tasks/s. Similarly, at 16 nodes, using 16 instances increases throughput from 43 to 195~tasks/s.

At larger node counts, increasing the number of instances continues to improve average task throughput, but with diminishing returns. For example, at 256 nodes, throughput rises modestly from 286.7 to 302.5 tasks/s when scaling from 1 to 64 instances. At 1024 nodes, throughput increases from 160.6 to 232.9 tasks/s as instances increase from 1 to 16---an improvement in percentage, but still lower than at 256 nodes. While average throughput growth flattens at scale, maximum throughput remains consistently high, reaching up to 930 tasks/s.

Importantly, task throughput increases not only with node count but also with instance count until reaching a saturation point. Resource utilization remains high ($\geq 94.5\%$) for all configurations up to 64 nodes. Beyond that, utilization begins to decrease due to the limits on concurrent \texttt{srun} invocations and the overhead of managing many Flux instances. For example, at 1024 nodes and 16 instances, utilization drops to 75.4\%.

Overall, this experiment shows that partitioning the resource allocation into multiple Flux instances is an effective strategy for increasing task throughput, especially for medium-scale jobs. The approach allows RP to flexibly balance scalability, performance, and fault isolation. However, its effectiveness above 1024 nodes is limited by the task launcher and runtime system initialization overheads, as well as system-imposed concurrency restrictions.

\subsubsection{Experiment \textit{Dragon}}

In this experiment, we measure the task launch throughput of a single Dragon instance across multiple nodes. Dragon is primarily optimized for executing in-memory Python functions, but it also supports executable tasks, similar to Flux. Therefore, we set up this experiment to launch executable tasks, using the same single-core sleep workload from previous experiments, to allow a direct comparison of the results.

Fig.~\ref{fig:all_throughput}(c) shows the average task throughput across three node counts: 4, 16, and 64. We observe that the average task throughput remains relatively stable between 4 and 16 nodes (343 and 380~tasks/s, respectively), but declines at 64 nodes to 204~tasks/s. Maximum throughput follows a similar trend, peaking at 622~tasks/s for 4 nodes and dropping to 272~tasks/s at 64 nodes.


Unlike Flux, where throughput scales with the number of nodes and instances, Dragon's task throughput appears largely independent of node count. This is expected: in this configuration, we use a single Dragon instance to manage all allocated nodes. The centralized design of Dragon in this experiment imposes scalability limits, particularly for workloads dominated by external process spawning rather than lightweight function execution.

These results highlight the current performance of Dragon when launching traditional executable workloads at scale. In later experiments (e.g., \textit{flux+dragon}), we evaluate Dragon in its native mode---executing Python functions---where we expect improved performance and scalability characteristics.

\subsubsection{Experiment \textit{Flux+\-Dra\-gon}}

In this experiment, we evaluate RP's ability to concurrently manage multiple runtime systems by running a mixed workload composed of both executable and Python function tasks. This configuration leverages RP's building-block architecture to integrate multiple instances of Flux and Dragon, using Flux to execute executable tasks and Dragon to execute in-memory Python function tasks.

This experiment measures the combined system's throughput and resource utilization, and assesses RP's ability to partition resources among multiple runtime systems and scale out each of them independently. Each configuration uses an equal number of Flux and Dragon instances, with each instance assigned a subset of the allocated nodes. Tasks are dispatched to the appropriate backend based on task type.

Fig.~\ref{fig:all_throughput}(d) shows the average task throughput on 2 to 64 nodes. Task throughput increases with the number of nodes and instances. At 16 nodes with 8 instances per runtime, the system achieves an average throughput of 171~tasks/s and a maximum of 573~tasks/s. At 64 nodes, maximum throughput peaks at 1547~tasks/s. This peak reflects the current upper bound of RP's task management subsystem, suggesting potential gains from further optimizing task dispatch performance within RP.


Resource utilization remains consistently high across all scales, with all configurations achieving at least 99.6\% and some reaching 100\%. This confirms the effectiveness of RP in managing hybrid runtimes and coordinating multiple backends for heterogeneous workloads.

Although the experiment uses equal resource partitioning for Flux and Dragon, RP supports independently configurable partition counts and sizes. This flexibility will enable future workloads to dynamically adjust the runtime ratio based on the task type distribution, allowing for workload-aware optimization strategies.

\begin{figure}
    \includegraphics[width=0.5\textwidth]{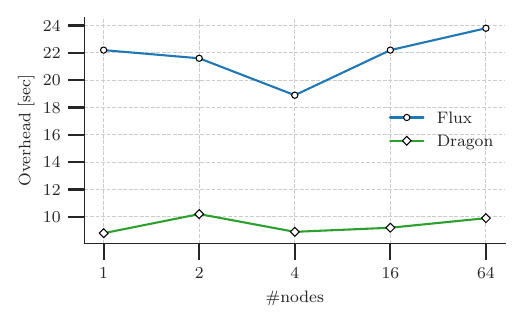}
    \caption{Instance launching overheads for an instance with 1 to 64 nodes. As instances launch concurrently, the overhead is not additive with respect to the number of instances.
    \label{fig:overheads}}
\end{figure}

Fig.~\ref{fig:overheads} shows the instance startup overhead for both Flux and Dragon. Overhead is defined as the time required to bootstrap each runtime system and prepare it to execute tasks. The results show that instance startup incurs a modest and relatively consistent cost---approximately 20 seconds for Flux and 9 seconds for Dragon---regardless of the amount of resources assigned to each instance. This suggests that overhead is dominated by runtime system initialization rather than instance scale. Since multiple instances are launched concurrently, the total overhead is not additive with respect to the number of instances.

Together, these results show that RP can efficiently manage heterogeneous workloads by concurrently orchestrating multiple runtime systems. The use of Flux and Dragon enables flexible execution strategies while maintaining high throughput and utilization at scale.

\subsection{Executing IMPECCABLE at Scale}\label{ssec:exp-usecase}

The IMPECCABLE use case, described in \S\ref{sec:uc}, requires the concurrent, asynchronous execution of multiple heterogeneous workflows with diverse resource requirements. Scaling such workflows on leadership-class platforms demands access to large numbers of CPU cores, GPUs, and sufficient memory, motivating our focus on Frontier, the first exascale HPC platform deployed in the USA.

Previous experiments show that RP’s default executor using \texttt{srun} on Frontier cannot meet IMPECCABLE’s scalability needs due to platform-imposed limits on concurrent task launching via Slurm. To address this, and consistent with the results of Experiment~\textit{flux\_1}, we developed a prototype implementation of IMPECCABLE that integrates RP with Flux.

We conducted two experiments to evaluate this prototype (see Table~\ref{tab:experiments}): \textit{im\-pec\-ca\-ble\_srun} and \textit{im\-pec\-ca\-ble\_flux}. Both use a dummy workload---sleep tasks---to isolate and compare the performance of RP with \texttt{srun} and Flux backends at scale. These controlled experiments allow us to assess backend-induced bottlenecks in task throughput, scheduling latency, and resource utilization.

All experiments introduce adaptive scheduling. Specifically, the number of tasks instantiated by some workflows is adjusted dynamically at runtime based on available system resources. This enables RP to opportunistically exploit idle compute resources, increasing overall system throughput. Additionally, RP monitors task failures and applies basic fault tolerance via retries or logging. For consistency across runs, we enforce a lower bound of 102 tasks per 128 nodes.

To further understand system behavior, we also measure the overhead introduced by RP and Flux when running null workloads. This includes the time taken to initialize, schedule, and place tasks onto allocated resources. Measuring this overhead provides insight into middleware efficiency and helps identify potential runtime bottlenecks.

\begin{figure}[ht]
    \centering
    \begin{subfigure}{\linewidth}
        \centering
        \includegraphics[width=\linewidth]{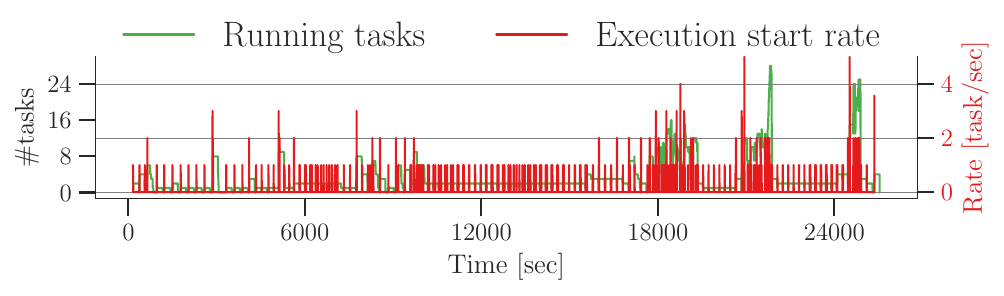}
        \caption{Srun backend, 256 nodes
                 \label{fig:srun256}}
    \end{subfigure}
    \vspace{1mm}
    \begin{subfigure}{\linewidth}
        \centering
        \includegraphics[width=\linewidth]{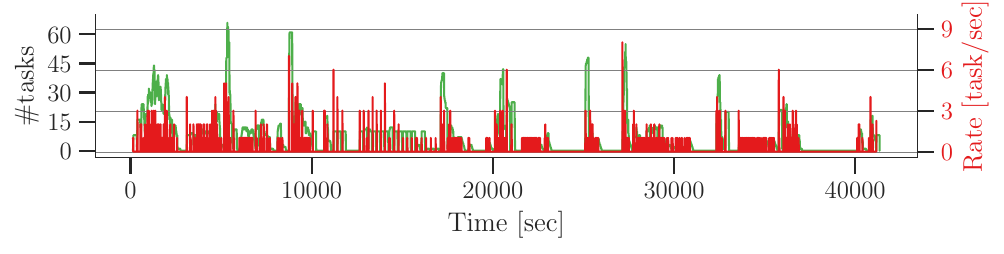}
        \caption{Srun backend, 1024 nodes\label{fig:srun1024}}
    \end{subfigure}
    \vspace{1mm}
    \begin{subfigure}{\linewidth}
        \centering
        \includegraphics[width=\linewidth]{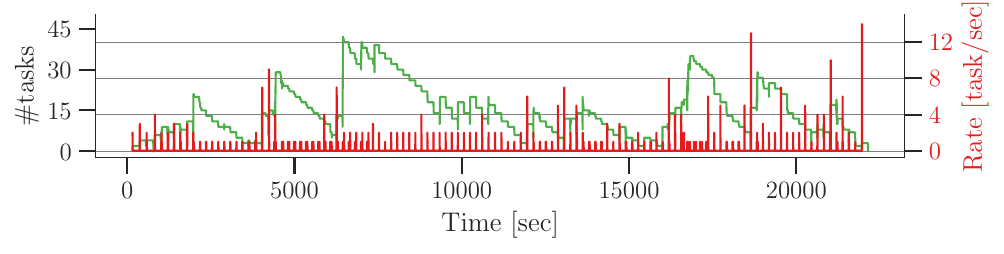}
        \caption{Flux backend, 256 nodes\label{fig:flux256}}
    \end{subfigure}
    \vspace{1mm}
    \begin{subfigure}{\linewidth}
        \centering
        \includegraphics[width=\linewidth]{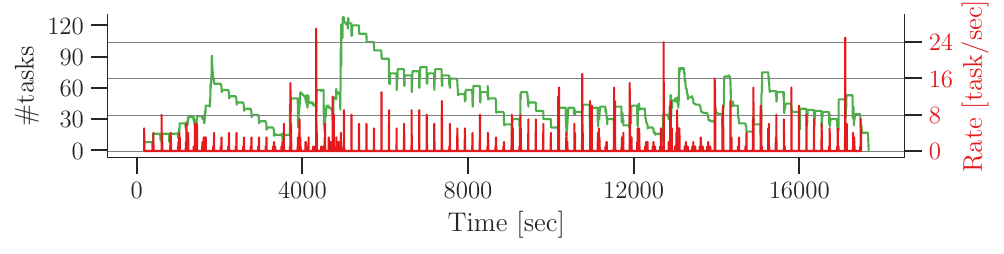}
        \caption{Flux backend, 1024 nodes\label{fig:flux1024}}
    \end{subfigure}
    \caption{Task concurrency (running green state, left y-axis) and execution start rate (red, right y-axis) for the IMPECCABLE run using a dummy workload (\texttt{sleep 180}). The runs use RADICAL-Pilot with \texttt{srun} (a,b) and Flux (c,d) as the execution backend over 256 and 1024 Frontier nodes, respectively.}
    \label{fig:null_impeccable}
\end{figure}

Fig.~\ref{fig:null_impeccable} presents the results of Experiments~\texttt{im\-pec\-ca\-ble\_\-srun} and~\texttt{im\-pec\-ca\-ble\_\-flux}, which compare RP’s execution behavior when using \texttt{srun} or Flux as the task launcher under otherwise identical workload configurations. Each configuration allocates either 256 or 1024 nodes and executes approximately 550 or 1800 heterogeneous tasks (using up to 7168 CPU cores and 1024 GPUs). All tasks sleep for 180 seconds, enabling isolation of key runtime metrics. Running tasks (green) represent concurrency, i.e., tasks actively executing. Execution start rate (red, right y-axis) denotes how quickly the backend launches tasks per second. Together, these plots expose the ability of RP and its execution backend to handle high-throughput, large-scale workloads.

The \texttt{srun} backend (Fig.~\ref{fig:srun256}, \ref{fig:srun1024}) consistently underperforms compared to Flux (Fig.~\ref{fig:flux256}, \ref{fig:flux1024}), especially in workflow makespan: approximately 26,000 and 44,000 seconds at 256 and 1024 nodes with \texttt{srun}, versus 22,000 and 17,500 seconds with Flux. These inflated makespans are attributable to the described platform-level limitations of \texttt{srun}.

Consistent with prior experiments, \texttt{srun}'s throughput (i.e., task launch rate) is both lower and more erratic, with frequent dips indicating contention or serialization in the launcher subsystem. This results in lower and more variable concurrency, as RP is forced to queue scheduled tasks while waiting for \texttt{srun} to become available. The consequence is underutilization of allocated resources: the number of running tasks consistently trails the number of scheduled tasks, with the gap widening at 1024 nodes.

In contrast, the Flux backend enables significantly more consistent and higher throughput. This results in tighter synchronization between scheduled and running tasks. Consequently, system concurrency is higher and more stable, leading to reduced idle time and improved resource utilization.

Resource utilization measures the fraction of allocated resources actively used by executing tasks. With the \texttt{srun} backend, CPU/GPU utilization reaches only 30\%/20\% at 256 nodes and drops to 15\%/14\% at 1024 nodes. In contrast, the Flux backend achieves 68\%/33\% utilization at 256 nodes and 69\%/43\% at 1024 nodes. Early experiments with the scientific workload show even higher efficiency: during the first four 12-hour steps at 256 nodes, CPU/GPU utilization reaches 98\%/30\%. Note that GPU utilization in this context depends on the scientific code, not on RP or Flux.

These results confirm that the \texttt{srun} backend imposes a significant bottleneck for executing IMPECCABLE via RP. Its inability to launch tasks promptly leads to a backlog in RP’s scheduler and inflated workflow makespans. In contrast, Flux removes these constraints, enabling RP to saturate available compute resources and reduce end-to-end execution time.

\subsection{Discussion and Impact}\label{ssec:exp-discussion}

The experimental results demonstrate that RP provides a flexible and scalable runtime system capable of adapting to diverse execution environments, workload types, and platform constraints. By decoupling task management from platform-native launchers and enabling backend-agnostic resource partitioning, RP consistently achieves high throughput and resource utilization across a broad range of runtime configurations.

Our baseline experiment with \texttt{srun} exposes a key limitation on Frontier: a system-level cap of 112 concurrent \texttt{srun} commands, which directly bounds task concurrency and prevents RP from fully utilizing available resources, even for modest workloads. This restriction highlights the need for more scalable launch mechanisms that can circumvent backend-imposed concurrency limitations.

Integrating Flux addresses this bottleneck. A single Flux instance (\textit{flux\_1}) sustains increasing throughput with job size, reaching over 700 tasks per second at 1024 nodes. When extended to multiple concurrent Flux partitions (\textit{flux\_n}), throughput and utilization improve further, although the gains taper off beyond 256 nodes due to coordination overheads and system-imposed limits on the number of concurrent \texttt{srun} invocations.

When Dragon launches executable tasks, it offers high through\-put at small and medium scales but exhibits limited scalability when deployed as a centralized backend across large node counts. However, its strengths become evident in hybrid configurations. The \textit{flux+dragon} experiment shows that RP can orchestrate multiple runtime systems concurrently, assigning tasks to the most suitable backend according to their implementation modality. This configuration achieves high launching throughput (up to 1,547~tasks/s) and near-perfect resource utilization, validating RP's runtime-agnostic architecture for managing heterogeneous workloads.

The IMPECCABLE experiments further contextualize these results. RP combined with Flux yields significantly shorter makespans, higher concurrency, and more stable execution behavior than the \texttt{srun}-based baseline, even under controlled dummy workloads. These improvements stem from RP's ability to partition resources, adaptively schedule tasks at runtime, and manage backend contention—capabilities essential for real-world workflows that require asynchronous, high-\allowbreak through\-put execution at scale.

Notably, the \textit{flux+dragon} experiment also lays the foundation for future implementations of IMPECCABLE that concurrently execute both HPC- and ML-style workloads. RP's ability to execute Python functions via Dragon and external executables via Flux enables task-type-aware backend selection. This hybrid capability not only sustains high throughput but also supports adaptive scheduling policies that reduce execution latency and improve workflow responsiveness. Such flexibility is critical in IMPECCABLE, where ML inference and HPC simulations are concurrently and asynchronously executed.  By demonstrating scalable, concurrent runtime integration, this experiment validates RP's architectural suitability for next-generation workflows that demand backend specialization and dynamic resource usage.

\section{Related Work}\label{sec:related}

Scientific workflow systems have long sought to enable scalable, efficient execution of complex applications on high-performance computing (HPC) platforms. Early systems such as Pegasus~\cite{deelman2019evolution}, Kepler~\cite{ludascher2006scientific}, and Swift/T~\cite{wozniak2013swift} offered structured workflow representations, data management, provenance capture, and basic scheduling. However, their reliance on external resource managers limits their ability to handle highly dynamic, asynchronous, and heterogeneous workloads on leadership-class platforms.

Traditional resource and job management systems (RJMSs) such as SLURM~\cite{yoo2003slurm}, PBS~\cite{jones2001pbs}, OAR~\cite{capit2005batch}, and LSF~\cite{zhou1992lsf} were primarily designed for large, tightly coupled MPI applications. Their limited support for fine-grained, concurrent task execution spurred the development of middleware tailored for many-task and ensemble workloads. Systems like TaskVine~\cite{delgado2023taskvine}, Parsl~\cite{babuji2019parsl}, Ray~\cite{moritz2018ray}, and Dask~\cite{rocklin2015dask} offer enhanced programmability and portability---often in cloud or hybrid settings---and have demonstrated strong scaling for specific workload types. However, they generally lack the deep integration with HPC runtimes needed to efficiently support heterogeneous and dynamically adaptive workloads at scale. Moreover, with the exception of Parsl, they do not implement the pilot abstraction that decouples resource acquisition from task execution, a capability critical for late binding and multi-level scheduling on leadership-class systems.

Pilot systems such as Glide\-in\-WMS~\cite{sfiligoi2009pilot}, Pan\-DA~\cite{maeno2024panda}, and DIRAC \cite{tsaregorodtsev2008dirac} implement multi-level scheduling and late binding, and have been deployed on HPC platforms. However, they are incompatible with the latest leadership-class systems and lack support for modern task runtimes such as Flux or Dragon. RADICAL-Pilot (RP)~\cite{turilli2018comprehensive} addresses this gap with a general-purpose pilot system for high concurrency, and heterogeneous tasks and resources. Its building-block architecture integrates with domain-specific workflow systems and runtime libraries, including RADICAL-AsyncFlow, Swift/T, Parsl, PanDA, and Apache Spark~\cite{asyncflow-docs}. This work extends earlier RP+Flux integration~\cite{lee2021scalable,turilli2024exaworks,al2021exaworks} by adding support for multiple concurrent Flux partitions, improving cross-partition scheduling, and implementing fault tolerance, increasing throughput from 14~tasks/s to 930~tasks/s.

The PMIx Reference RunTime Environment (PRRTE)~\cite{elwasif2023supporting} offers a lightweight, open-source runtime for scalable task launching. Unlike Flux, PRRTE does not include an internal scheduler but instead delegates coordination and scheduling to external systems. Its distributed virtual machine (DVM) model enables rapid task launch with minimal per-task overhead, provided task coordination is managed externally. Our work demonstrated how RP complements PRRTE’s minimalist design by supplying scheduling, fault tolerance, and coordination logic~\cite{titov2022radical}. To our knowledge, RP is the only pilot system that has integrated and evaluated PRRTE as a backend runtime at scale on leadership-class machines.

Flux~\cite{ahn2020flux} represents a new class of hierarchical, composable RJMSs designed to support scalable, dynamic scheduling on exascale systems. By providing APIs for task co-scheduling, fault tolerance, and resource partitioning, Flux enables capabilities traditionally found in pilot systems. When integrated with RP, Flux enables a hybrid execution model that, ultimately, will combine hierarchical scheduling with task-level adaptivity and runtime introspection.

Dragon~\cite{dragonhpc2025} occupies a distinct point in the design space: it is a high-throughput launcher designed to minimize overhead and coordination costs. Unlike PRRTE and Flux, Dragon does not support hierarchical composition, internal scheduling, or resource partitioning. However, it achieves unmatched launch throughput for short-lived, homogeneous function tasks, while offering task- and runtime-side abstractions for multi-node shared memory. When paired with RP, Dragon enables a design point where RP assumes full control over scheduling and coordination, and Dragon maximizes launch efficiency. Similar approaches were seen in tightly coupled systems like ALPS~\cite{gayen2012alps} and IBM’s jsrun~\cite{jsrun-ibm}, but those were tied to specific vendor stacks and lacked composability.

In prior work~\cite{turilli2019characterizing}, we characterized the performance of RP when integrated with ORTE (a predecessor of PRRTE) and JSM on Summit. That study established foundational results on scalability and resource utilization for many-task workloads. The present work significantly extends that baseline by demonstrating, for the first time, an integrated performance evaluation of RP with Flux and Dragon. We show how RP’s building-block architecture enables tailored and concurrent integration with each runtime. Through these integrations, RP becomes the first runtime system to support hierarchical scheduling, adaptive task execution, and high-throughput launching within a single extensible framework, supporting exascale systems such as Frontier.

\section{Conclusions}\label{sec:conclusion}

This paper shows how RADICAL-Pilot (RP), extended with Flux and Dragon, enables the execution of heterogeneous workflows on the leadership-class HPC platform Frontier at scale. Integrating multiple runtime systems within RP’s building-block architecture addresses a central challenge in modern scientific computing: no single runtime can efficiently and effectively support the diverse execution models of real-world workflows like IMPECCABLE.

We show that RP can coordinate the concurrent execution of heterogeneous tasks---including Python functions, external executables, and tightly coupled jobs---within a single pilot allocation. Flux provides hierarchical scheduling and resource partitioning, enabling fine-grained control over concurrency and throughput. Dragon offers lightweight task launching with high throughput, particularly suited to in-memory function execution. Combined, these backends enable RP to match each task type to the most appropriate runtime system, avoiding the scalability bottlenecks of traditional launchers such as \texttt{srun}.

Our experiments characterize RP’s performance across three key metrics: task throughput, resource utilization, and runtime overhead. Flux achieves over 700~tasks/s with a single instance and up to 930~tasks/s when using multiple concurrent partitions. In hybrid configurations with both Flux and Dragon, RP reaches over 1,500~tasks/s while maintaining 99.6--100\% resource utilization. When applied to the IMPECCABLE.v2 campaign, RP with Flux reduces workflow makespan by 30--60\% and sustains high concurrency and scheduling rates on up to 1024 nodes, even under dynamic and adaptive task loads.

In contrast, \texttt{srun}-based launching is limited by system-imposed concurrency ceilings, which cap achievable throughput and resource utilization. While \texttt{srun} peaks at 152~tasks/s on a single node, performance rapidly degrades to 61~tasks/s at 4 nodes, and continues to decline at larger scales. This constraint leads to persistent underutilization of available resources, highlighting the need for alternative backends to support high-throughput, asynchronous workflows at scale.

This work constitutes the first integrated performance study of RP with both Flux and Dragon, establishing a new execution model for extreme-scale workflows. By decoupling resource acquisition from task execution and enabling concurrent control over multiple runtime backends, RP provides a flexible and performant foundation for next-generation workflows that combine HPC simulation, ML inference, and asynchronous control.

Future work will explore tighter coupling across runtime systems, support for additional execution models (e.g., streaming or interactive agents), and dynamic backend selection based on workload characteristics. RP's extensibility offers a clear path forward for supporting emerging scientific use cases that demand both scale and diversity in execution models.

\begin{acks}

\footnotesize{A. Merzky, M. Titov and M. Turilli equally contributed to this paper. This work is supported in part by the following grants: NSF-2103986 and 1931512, and US DOE DE-AC02-06CH11357 (LUCID). We thank Agastya Bhati and Peter Coveney for insights and discussions on IMPECCABLE workloads.} {{\bf Experiments} Data and analysis scripts can be found at: {\footnotesize \url{https://github.com/radical-experiments/rp-flux-dragon}}}

\end{acks}

\small
\bibliographystyle{ACM-Reference-Format}
\bibliography{references.bib}

\end{document}